\documentclass[conference]{IEEEtran}
\IEEEoverridecommandlockouts
\usepackage{amsmath,amsfonts,amssymb}
\usepackage{graphicx, subcaption}
\usepackage{cite, authblk}

\title{Adaptive Calibration of RF Digital Twins via Budget-Constrained Radio Measurement Selection}

\author{Yalin E. Sagduyu, Tugba Erpek, and Hossein Jafari}
\affil{\normalsize Nexcepta, Gaithersburg, MD, USA}

\IEEEoverridecommandlockouts

\begin{document}
\maketitle
\thispagestyle{empty}

\begin{abstract}
Radio-frequency (RF) digital twins (DTs) enable site-specific wireless propagation prediction by explicitly reconstructing the physical environment and modeling electromagnetic interactions. Although physics-based DTs reduce reliance on exhaustive measurements, systematic bias relative to real radio data arises from environmental uncertainty, material inaccuracies, and geometric abstractions. Calibration using field measurements is therefore necessary but costly under limited sampling budgets. This paper studies active RF DT calibration, in which measurement locations are selected sequentially and in an online manner to estimate a calibration offset under strict measurement budget constraints. A physics-based RF DT pipeline is developed, integrating semantic geospatial data from OpenStreetMap, 3D modeling in Blender, and deterministic ray tracing in Sionna to capture dominant building and foliage effects. This RF DT is calibrated using path-loss measurements collected with software-defined radios. Multiple online calibration strategies that select radio measurement locations are compared, including random sampling, rank-structured sampling with balanced coverage, quasi-Monte Carlo ranked selection, progressive structure-aware allocation, and fully adaptive uncertainty-driven sampling with estimator stabilization. Results demonstrate that adaptive and structure-aware schemes significantly improve both mean and worst-case calibration accuracy, particularly in the low-budget regime for radio measurements.
\end{abstract}


\section{Introduction}
\emph{Physics-based radio-frequency (RF) digital twins (DTs)} enable site-specific modeling of wireless propagation in complex environments by explicitly reconstructing physical geometry and electromagnetic (EM) interactions. They provide path-loss and received-power predictions at arbitrary locations without exhaustive field measurements \cite{khan2022digital}. Such capabilities are critical for the design and optimization of modern wireless systems, especially in dense or irregular environments where empirical models fail to capture site-specific effects \cite{belgiovine2025better}.

In practice, RF DTs are built through a \emph{multi-stage} pipeline that converts geographic data into a simulation-ready EM scene \cite{testolina2024boston, lee2025analysis, zahid2025real, amatare2024real}. Despite their interpretability and spatial resolution, the fidelity of RF DTs is limited by geospatial uncertainty, geometric simplifications, and approximate material parameters, leading to systematic bias relative to real radio measurements. Even when spatial trends are well captured, a dominant bias can significantly degrade prediction accuracy. \emph{Calibration} of RF DTs using real measurements is therefore necessary \cite{ruah2026bridge, lai2023raytracing, ruah2024calibrating}, preserving spatial structure while compensating for aggregate bias. However, because radio measurements are \emph{expensive and time-consuming}, DT calibration must operate under strict radio measurement-budget constraints.

This motivates the problem of \emph{active DT calibration}, where measurement locations are selected sequentially to maximize calibration accuracy under a bounded measurement budget. Unlike passive approaches based on fixed or random placement, active calibration leverages DT predictions and observed discrepancies to guide future measurements. We study this problem for a physics-based RF DT constructed from OpenStreetMap (OSM) data, Blender-based 3D modeling, and Sionna-based deterministic ray tracing \cite{hoydis2023sionna}. Geospatial data are translated into explicit 3D geometry, EM materials are assigned, and ray tracing resolves line-of-sight and multipath propagation, yielding physics-grounded link-level predictions.

This RF DT is calibrated with \emph{real RF measurements} that we have collected using software-defined radios (SDRs). The objective is to minimize the offset between field-measured and DT-simulated path losses. For active DT calibration, we compare five schemes for selecting measurement locations under a strict budget: (i) Random Sampling (RS); (ii) Balanced Rank Sampling (BRS), an online rank-structured scheme that promotes balanced coverage of the DT prediction range through symmetric rank pairing; (iii) Quasi-Monte-Carlo Ranked Sampling (QRS), which enforces systematic coverage of the ranked DT space using a low-discrepancy sequence; (iv) Progressive Structure-Aware Sampling (PSS), which uses DT prediction variability with progressive adaptive allocation, stable late-stage behavior, and robust estimation near budget exhaustion; and (v) Adaptive Uncertainty-Aware Sampling (AUS), which adapts allocation using residual-informed uncertainty proxies and stabilizes offset estimation via exponential smoothing.

We show a consistent performance ordering across budgets: RS yields the largest mean and worst-case errors; BRS improves early robustness via balanced rank coverage; QRS strengthens coverage efficiency; PSS enhances robustness and convergence through progressive, stabilized allocation; and AUS achieves the fastest error decay and smallest worst-case error, especially in the low-budget regime. Overall, adaptive and structure-aware schemes can significantly increase calibration accuracy with a limited number of radio measurements. 

The remainder of the paper is organized as follows. Section~II presents the RF DT construction pipeline. Section~III formulates the DT calibration problem under measurement constraints. Section~IV details the online schemes. Section~V reports numerical results. Section~VI concludes the paper.

\section{Physics-Based RF Digital Twin Construction}
\begin{figure}[t!]
  \centering
  \includegraphics[width=\columnwidth]{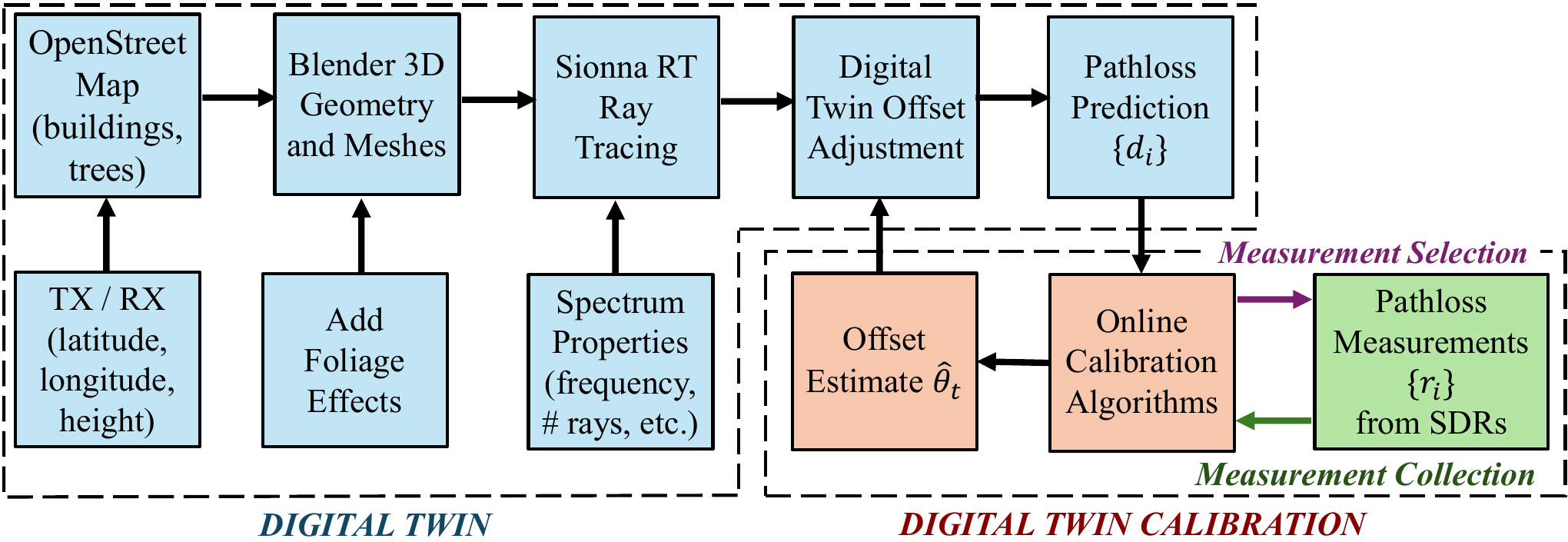}
  \caption{RF DT pipeline with active calibration.}
  \label{fig:DT_system} \vspace{-0.35cm}
\end{figure}

The RF DT is constructed as a physics-aware representation of the radio environment by combining semantic geospatial data from OSM using actual location coordinates as the input, 3D modeling in Blender, and deterministic propagation via the Sionna ray-tracing framework, as shown in Fig.~\ref{fig:DT_system}. This RF DT construction pipeline converts geographic data into a simulation-ready scene that captures geometry, material properties, and radio-wave interactions.

\noindent \textbf{Spatial Reference and Local Coordinate System.}
Transmitter and receiver locations are specified in geodetic latitude-longitude and projected into a locally valid Cartesian frame in meters. The projected coordinate system is selected automatically using the geographic midpoint to minimize distortion, and a local tangent frame is defined by translating coordinates so the origin lies near the scene center. All geometry (buildings, vegetation, and radio nodes) is expressed in this frame for numerical robustness and physical consistency.

\noindent \textbf{Semantic Environment Acquisition from OSM.} The environment is populated using semantic geospatial data from OSM via the Overpass API. Building footprints are retrieved as polygons, while vegetation is extracted as point features representing individual trees or rows. Semantic tags are later mapped to geometric abstractions and EM material models.

\begin{figure}[t]
    \centering
    \begin{subfigure}{0.675\linewidth}
        \centering
        \includegraphics[width=\linewidth]{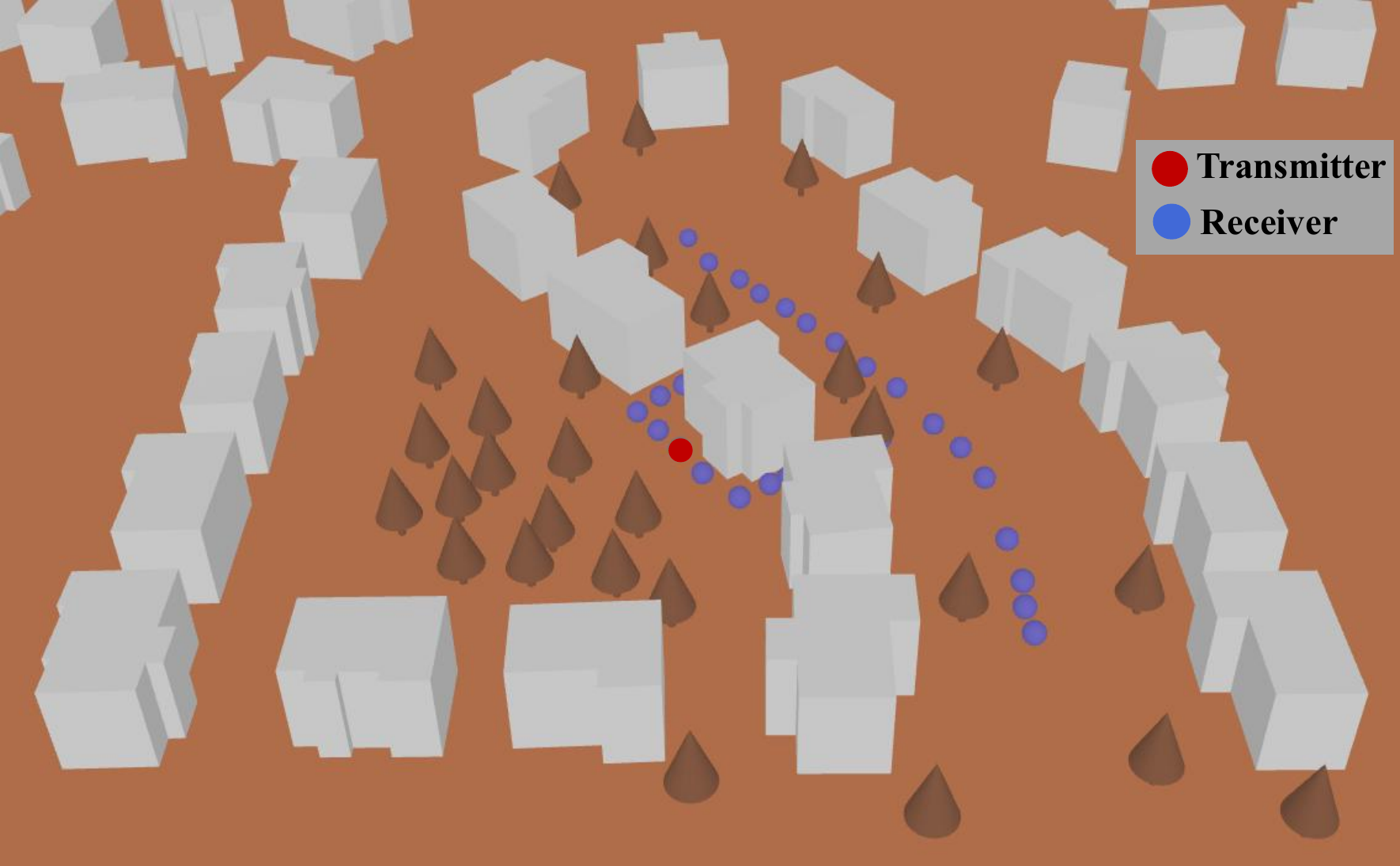}
        \caption{Oblique view.}
        \label{fig:Blender_oblique}
         \vspace{0.2cm}
    \end{subfigure}
   
    \begin{subfigure}{0.675\linewidth}
        \centering
        \includegraphics[width=\linewidth]{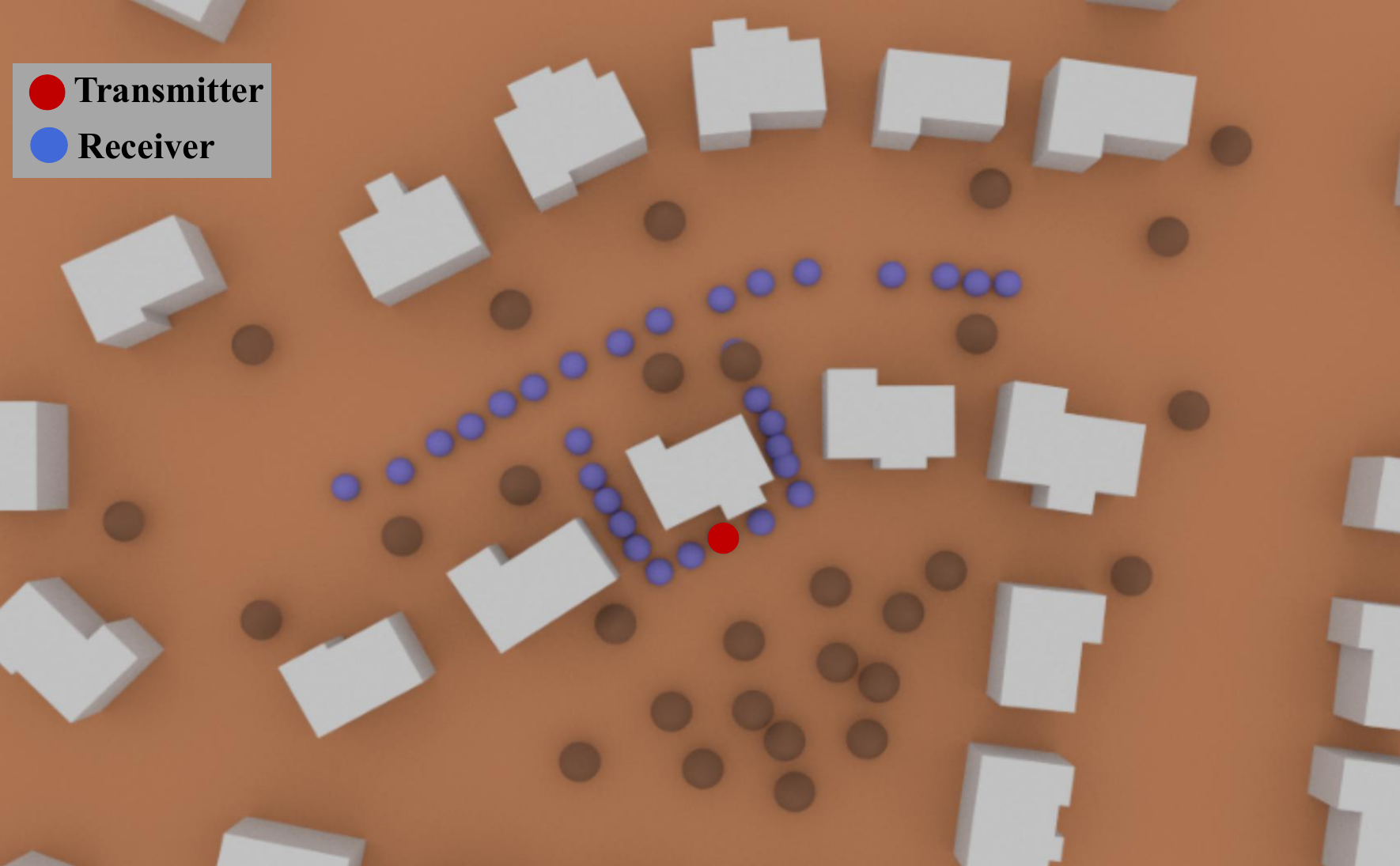}
        \caption{Top view.}
        \label{fig:Blender_top}
    \end{subfigure}
    \caption{Blender scene for the RF DT.}
    \label{fig:Blender} \vspace{-0.45cm}
\end{figure}

\noindent \textbf{Geometric Modeling in Blender.}
Geospatial data are converted into explicit 3D geometry in Blender. Building footprints are vertically extruded into volumetric solids, a continuous ground surface is added, and vegetation is represented using simplified trunk-crown proxies. This abstraction captures dominant EM effects (blockage, attenuation, and scattering) while preserving manageable complexity for deterministic ray tracing. Beyond OSM-derived vegetation, the DT supports augmentation to compensate for missing features and enable controlled propagation studies. Additional trees are specified in the local Cartesian frame and injected as explicit geometric objects. Each injected tree consists of a vertical cylindrical trunk and a cone-shaped crown. Trunk and crown parameters are varied to match the actual vegetation size and density. The resulting Blender scene is shown in Fig.~\ref{fig:Blender}.

\noindent \textbf{Mesh Conditioning and Scene Assembly.}
All geometric objects are conditioned for deterministic ray tracing through triangulation with consistent surface normals and exported as individual mesh assets. These meshes are assembled into a unified scene description encoding spatial relationships and object transforms, thereby decoupling geometric modeling from EM simulation. Vegetation primitives generated in Blender integrate directly into this pipeline.

\noindent \textbf{Material Semantics and EM Properties.}
Each object class is assigned a frequency-dependent EM material model. Ground, buildings, trunks, and crowns are modeled with distinct dielectric and conductive properties, with trunks and foliage treated separately to reflect their differing EM behavior. This separation enables systematic evaluation of the impact of geometry and material assumptions on propagation predictions.

\noindent \textbf{Ray-Tracing Propagation Modeling.}
The assembled DT is instantiated in the Sionna ray-tracing framework. Transmitters and receivers are placed in the local coordinate frame, and deterministic ray tracing enumerates multipath components from line-of-sight, reflection, diffraction, scattering, and refraction. Each path is characterized by complex gain and delay, forming a physics-grounded channel model. The scene is exported in Mitsuba-compatible format, enabling Sionna to utilize Mitsuba's physically based rendering and ray-geometry intersection engine for deterministic propagation modeling. Additionally, injected vegetation, represented as explicit mesh geometry with consistent material assignments, is treated identically to OSM-derived objects, enabling direct assessment of its impact on multipath structure and path loss.

\noindent \textbf{Link-Level Metrics.}
Received power and path loss are computed by aggregating all resolved propagation paths and scaling by transmit power. These metrics directly reflect the encoded geometry and material properties. Local vegetation augmentation provides a practical means to mitigate bias from incomplete environmental data and to systematically evaluate sensitivity of link metrics to vegetation placement and density.

\section{Digital Twin Calibration}
The pipeline for active DT calibration is shown in Fig.~\ref{fig:DT_system}.
Consider a set of $N$ candidate spatial locations indexed by
$\mathcal{L}=\{1,\dots,N\}$. At each location $i\in\mathcal{L}$, the DT provides a
predicted large-scale path-loss value $d_i$. The true path-loss $r_i$ at location
$i$ can only be obtained via a real radio measurement. 

The residual
between the DT prediction and the true measurement at location $i$ is given by
\begin{equation}
e_i = d_i - r_i .
\end{equation}
We calibrate the DT using a global offset correction model, where the calibrated
DT prediction is
\begin{equation}
\tilde d_i = d_i - \theta ,
\end{equation}
and $\theta$ is a scalar offset. The optimal (oracle) calibration offset is the
mean residual over all locations,
\begin{equation}
\theta^\star = \frac{1}{N}\sum_{i=1}^N e_i
= \frac{1}{N}\sum_{i=1}^N (d_i - r_i). \label{eq:oracle_offset}
\end{equation}
The oracle offset is used for evaluation and is not available to the online schemes. Due to practical constraints, only $k \leq N$ radio measurements are collected. Let $\mathcal{M}_k \subset \mathcal{L}$ denote the set of $k$ locations at which $k$ measurements are performed. An estimator $\hat{\theta}_k$ of the calibration offset is computed using the residuals observed at these locations.
Active DT calibration selects measurement locations such that $\hat\theta_k$ closely approximates $\theta^\star$ using only $k$ measurements.

To quantify DT calibration fidelity, we define the absolute offset error after $k$ measurements in Monte Carlo run  $j$ as
\begin{equation}
\Delta_{k,j} \triangleq \big| \hat{\theta}_{k}^{(j)} - \theta^\star \big|,
\end{equation}
where $\hat{\theta}_{k}^{(j)}$ denotes the offset estimate produced by a given scheme in run $j$ using $k$ measurements. To account for randomness, each experiment is repeated over $T$ randomized runs (e.g., different random seeds and randomized shifts in rank-based selection). Performance is summarized across runs using the mean and worst-case errors,
\begin{align}
\Delta_k^{\mathrm{mean}} &\triangleq \frac{1}{T}\sum_{j=1}^T \Delta_{k,j}, \nonumber\\
\Delta_k^{\max}          &\triangleq \max_{j \in \{1,\dots,T\}} \Delta_{k,j}.
\label{eq:metrics}
\end{align}
which capture typical calibration accuracy and worst-case deviation from the
oracle offset, respectively. 

\section{Active Digital Twin Calibration Schemes}
A real RF measurement from one location is collected at each step. Let $\mathcal{M}_t \subset \mathcal{L}$ denote the set of locations measured after $t$ steps, with $|\mathcal{M}_t| = t$. At online step
$t \in \{1,\dots,k\}$, the calibration offset estimate is denoted by
$\hat{\theta}_t$. A DT measurement selection scheme selects location $
i_t \in \mathcal{L} \setminus \mathcal{M}_{t-1}$,
observes $r_{i_t}$, forms $e_{i_t}=d_{i_t}-r_{i_t}$, and updates an offset estimate $\hat{\theta}_t$. All schemes are \emph{online}: $i_t$ depends only on $\{d_i\}_{i=1}^N$ and previously observed residuals $\{e_i : i \in \mathcal{M}_{t-1}\}$. Selection and estimation are jointly designed per scheme. 

\subsection{Random Sampling (RS)}

RS draws $i_t$ uniformly without replacement from $\mathcal{L}\setminus\mathcal{M}_{t-1}$. The offset estimate is the empirical mean
\begin{equation}
\hat{\theta}_t^{\mathrm{mean}}
= \frac{1}{t}\sum_{i\in\mathcal{M}_t} e_i .
\end{equation}
Under uniform sampling without replacement, the empirical mean is unbiased for $\theta^\star$. For small $t$, variance may be large in heterogeneous environments. BRS and QRS use the same estimator but differ in selection.

\subsection{Balanced Rank Sampling (BRS)}

Let $\rho(i)\in\{0,\dots,N-1\}$ denote the rank of $d_i$ in ascending order. Target ranks are generated using a randomly shifted low-discrepancy sequence to ensure uniform coverage of the ordered DT space. Define the set of unmeasured ranks
\begin{equation}
\mathcal{R}_{t-1}
=
\{\rho(i): i\in\mathcal{L}\setminus\mathcal{M}_{t-1}\}.
\end{equation}

At step $t$, BRS sets target rank $r_t=\lfloor N u_t\rfloor$, where $u_t\in[0,1)$ is the $t$-th shifted low-discrepancy sample, and selects 
\begin{equation}
i_t=\arg\min_{j\in\mathcal{L}\setminus\mathcal{M}_{t-1}} |\rho(j)-r_t|
\end{equation} 
as the nearest unmeasured rank. BRS uses rank reflection:
\begin{equation}
\rho_t^{\mathrm{pair}} =
\begin{cases}
(N-1)-\rho_t, & \text{if } (N-1)-\rho_t \in \mathcal{R}_{t-1},\\
\varnothing, & \text{otherwise},
\end{cases}
\end{equation}
where $\rho_t=\rho(i_t)$ is the rank selected at step $t$. If feasible and budget permits, the paired location is also measured. Near budget exhaustion, structure is disabled and selection reverts to uniform sampling. BRS is structured but non-adaptive.

\subsection{Quasi-Monte-Carlo Ranked Sampling (QRS)}
With $\rho(i)$ defined as above and $r_t=\lfloor N u_t\rfloor$, QRS selects
\begin{equation}
i_t=\arg\min_{j\in\mathcal{L}\setminus\mathcal{M}_{t-1}}
|\rho(j)-r_t|,
\end{equation}
to enforce systematic coverage of the ordered DT prediction range without explicit rank-reflection pairing. Uniform coverage arises naturally from the low-discrepancy sequence itself. As in RS and BRS, estimation uses $\hat{\theta}_t^{\mathrm{mean}}$. QRS also reverts to uniform selection when few candidates remain.

\subsection{Progressive Structure-Aware Sampling (PSS)}

PSS exploits dispersion in DT predictions while ensuring tail stability. The index set is partitioned into $B$ bins $\{\mathcal{L}_b\}$ via empirical quantiles, with $N_b=|\mathcal{L}_b|$ and $\pi_b=N_b/N$. Within-bin dispersion is quantified by the interquartile range
\begin{equation}
\Delta_b =
Q_{0.75}(\{d_i:i\in\mathcal{L}_b\})
-
Q_{0.25}(\{d_i:i\in\mathcal{L}_b\}),
\end{equation}
where $Q_q(\cdot)$ denotes the empirical $q$-quantile. A target allocation balancing coverage and dispersion is
\begin{equation}
a_b =
(1-\eta)\pi_b
+
\eta
\frac{N_b \Delta_b}{\sum_{b'} N_{b'} \Delta_{b'}},
\end{equation}
where $\eta\in(0,1)$. At each step, the bin with largest allocation deficit is selected. Within that bin, the next point maximizes DT-space diversity relative to previously selected points in the same bin. When the remaining candidate set becomes small, PSS falls back to uniform random selection.

For estimation, PSS uses a shrinkage-stratified estimator. Let $\bar e(t)=\frac{1}{t}\sum_{i\in\mathcal{M}_t} e_i$ be the global mean residual, and $\bar e_b(t)$ the within-bin mean with $m_b(t)$ measured points. Define
\begin{equation}
\tilde e_b(t)=
\begin{cases}
\dfrac{m_b(t)\bar e_b(t)+\kappa\,\bar e(t)}{m_b(t)+\kappa}, & m_b(t)>0,\\
\bar e(t), & m_b(t)=0,
\end{cases}
\end{equation}
where $\kappa \! > \!0$ controls mean shrinkage. The stratified estimate
\begin{equation}
\hat\theta_t^{\mathrm{strat}}=\sum_{b=1}^B \pi_b\,\tilde e_b(t).
\end{equation}
is formed from bin estimates. To prevent late-budget artifacts,
\begin{equation}
\hat\theta_t^{\mathrm{PSS}} = w_t\,\hat\theta_t^{\mathrm{strat}} + (1-w_t)\,\hat\theta_t^{\mathrm{mean}},
\end{equation}
where $w_t$ decreases toward zero as $t$ approaches $k$. 

\subsection{Adaptive Uncertainty-Aware Sampling (AUS)}

AUS combines residual-informed uncertainty allocation with estimator stabilization. Using the same bin partition, let
\[
m_b(t) = |\mathcal{M}_t \cap \mathcal{L}_b|.
\]

For $m_b(t)\ge 2$, the within-bin variance is estimated as 
\begin{equation}
\hat{\sigma}_b^2(t) =
\frac{1}{m_b(t)-1}
\sum_{i \in \mathcal{M}_t \cap \mathcal{L}_b}
\big(e_i - \bar e_b(t)\big)^2 ,
\end{equation}
else $\hat{\sigma}_b^2(t)=0$. The global variance is
\begin{equation}
\hat{\sigma}_g^2(t) =
\frac{1}{t-1}
\sum_{i \in \mathcal{M}_t}
\big(e_i - \hat{\theta}_t^{\mathrm{mean}}\big)^2 .
\end{equation}

Variance shrinkage stabilizes bin-level estimates:
\begin{equation}
\sigma_b^2(t) =
\frac{m_b(t)}{m_b(t)+\tau}\hat{\sigma}_b^2(t)
+
\frac{\tau}{m_b(t)+\tau}\hat{\sigma}_g^2(t),
\end{equation}
with $\tau>0$. For $p\in(0,1]$ and $\epsilon>0$, bin scores are
\begin{equation}
w_b(t) = \big(\sigma_b^2(t)\big)^{p/2} + \epsilon,
\end{equation}
Then coverage and variance allocations are combined as
\begin{equation}
a_b(t)
=
(1-\alpha_t)\pi_b
+
\alpha_t
\frac{N_b w_b(t)}{\sum_{b'} N_{b'} w_{b'}(t)},
\end{equation}
with increasing $\alpha_t$. The next bin is chosen via
\begin{equation}
b_t
=
\arg\max_b
\big(a_b(t)\,t - m_b(t-1)\big).
\end{equation}

Within bin $b_t$, the selected index maximizes DT diversity,
\begin{equation}
i_t
=
\arg\max_{j \in \mathcal{L}_{b_t}\setminus\mathcal{M}_{t-1}}
\min_{i \in \mathcal{M}_{t-1}\cap\mathcal{L}_{b_t}}
|d_j - d_i|.
\end{equation}
If no prior measurement exists in $b_t$, the unmeasured point closest to the bin median is selected. The raw estimate equals $\hat{\theta}_t^{\mathrm{mean}}$. To stabilize estimation under limited budgets, AUS applies Exponentially Weighted Moving Average (EWMA):
\begin{equation}
\hat{\theta}_t^{\mathrm{AUS}}
=
(1-\lambda_t)\hat{\theta}_{t-1}^{\mathrm{AUS}}
+
\lambda_t \hat{\theta}_t^{\mathrm{mean}},
\end{equation}
where $\lambda_t$ follows a capped-decay schedule (e.g., $\lambda_t=\min(\lambda_{\max}, c/t)$ with a positive late-stage floor). The estimator is initialized with $\hat{\theta}_1^{\mathrm{AUS}} = \hat{\theta}_1^{\mathrm{mean}}$.

\subsection{Complexity}
In all schemes, computational cost is negligible relative to DT ray tracing and radio measurements. Computational (time) complexity is $\mathcal{O}(k)$ for RS. BRS and QRS require an initial sorting step of $\mathcal{O}(N\log N)$ to obtain ranks, after which selection is $\mathcal{O}(k)$. PSS and AUS additionally perform bin-based allocation and within-bin diversity search; their per-step cost is $\mathcal{O}(B + N_b)$, where $B$ is the number of bins and $N_b$ the bin size, leading to worst-case $\mathcal{O}(kN/B)$ online complexity. In the worst case where bins are highly unbalanced, the online complexity may approach $\mathcal{O}(kN)$.

\section{Performance Evaluation}
We collected RF path-loss measurements using USRP B210 radios with a transmit power of $10$~dBm at $30$ receiver locations ($N=30$) and performed DT simulations at the same locations. Fig.~\ref{fig:Blender} shows the DT scene generated from the measurement location coordinates. The carrier frequency was set to $1.2\,\mathrm{GHz}$ for both the DT and radio experiments. In the DT, the maximum ray interaction depth was set to $5$, with $200{,}000$ rays launched per source, and material assignments followed ITU-based models for buildings, ground, and trees. 

\begin{figure}[t!]
  \centering
  \includegraphics[width=0.8\columnwidth]{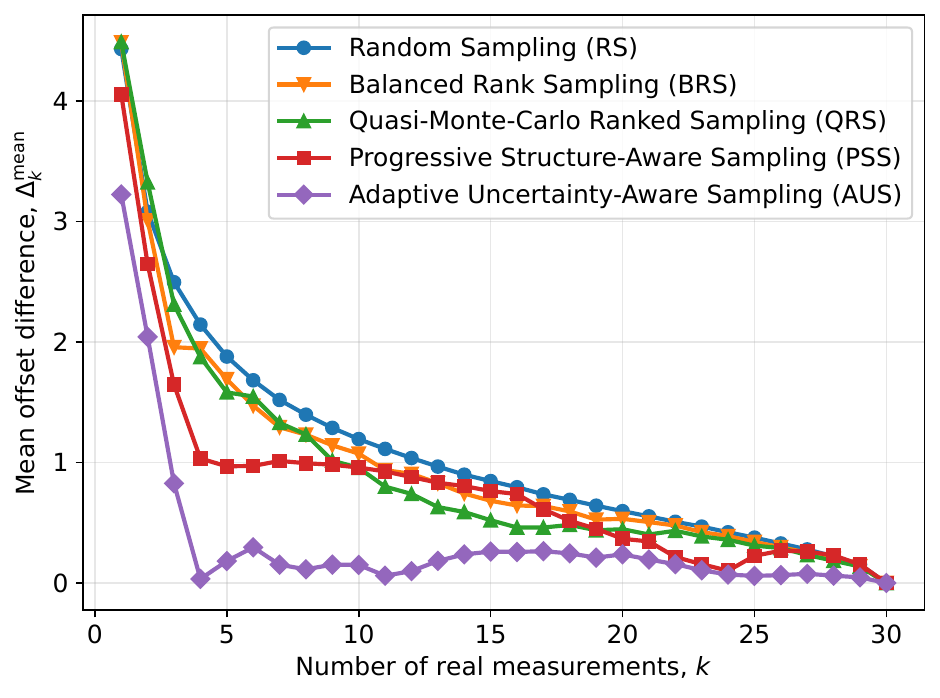}
  \caption{Mean absolute offset error in RF DT calibration.}
  \label{fig:mean} 
\vspace{0.35cm}
  \centering
  \includegraphics[width=0.8\columnwidth]{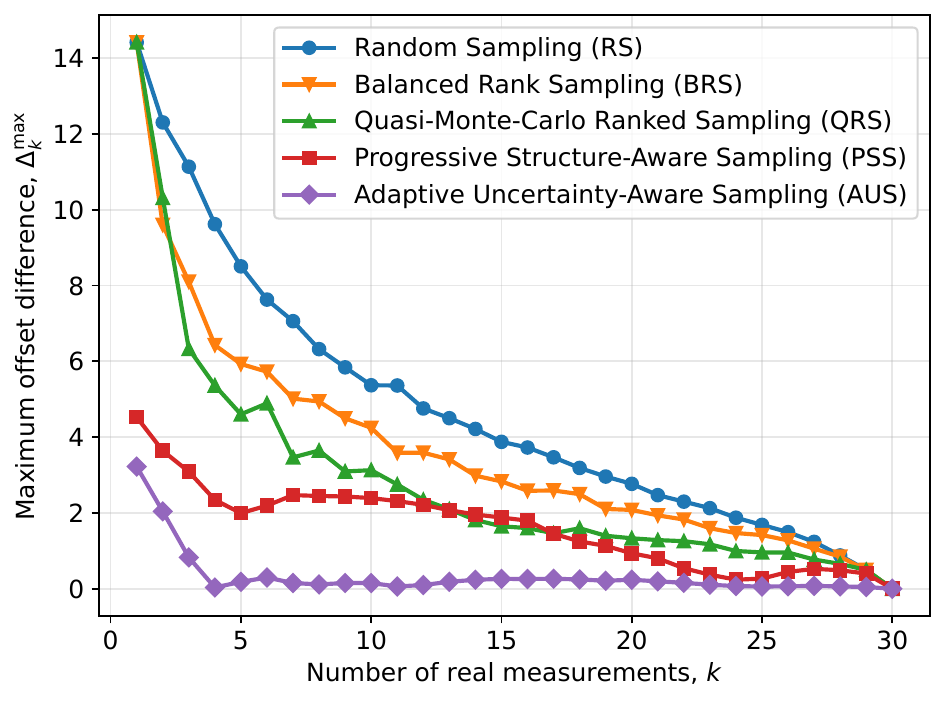}
  \caption{Worst-case absolute offset error in RF DT calibration.}
  \label{fig:max} \vspace{-0.15cm}
\end{figure}

The numerical evaluation of DT calibration follows an offline Monte Carlo procedure that emulates online active calibration under strict measurement-budget constraints. A fixed set of $N$ DT predictions $\{d_i\}$ and corresponding ground-truth radio measurements $\{r_i\}$ is used to compute the true global offset defined in (\ref{eq:oracle_offset}). For each calibration scheme and each measurement budget $k \in \{1,\dots,N\}$, the online selection and estimation procedure is executed sequentially for $k$ steps, with residuals revealed only at selected locations. Each experiment is repeated over multiple randomized runs (different random seeds and randomized shifts in rank-based selection). Performance is measured using the mean and worst-case errors defined in (\ref{eq:metrics}) and shown in Figs.~\ref{fig:mean}-\ref{fig:max} as a function of the measurement budget $k$. 

The error generally decreases with $k$, although minor local fluctuations may occur because each newly measured path-loss residual $e_i=d_i-r_i$ introduces discrete variability before the growing measurement set stabilizes the estimate. For very small $k$, RS exhibits large mean error and severe worst-case outliers, reflecting high-variance estimation. Rank-structured strategies reduce both metrics even with few measurements. BRS improves over RS by promoting balanced coverage of the DT prediction range through symmetric rank pairing in the ordered DT space, thereby reducing the probability of highly unrepresentative early selections. QRS further improves performance by enforcing more uniform global coverage of the ranked DT axis through low-discrepancy placement, yielding smaller mean and worst-case errors across most budgets.

PSS provides additional gains by coupling DT-only dispersion-driven allocation with a progressive deficit rule and late-stage stabilization. This improves intermediate-regime robustness and mitigates late-budget artifacts, resulting in more reliable convergence than purely rank-coverage schemes. AUS achieves the fastest error decay for small $k$ and the smallest worst-case deviations by combining residual-informed adaptive allocation with EWMA stabilization, which accelerates learning while damping early noise.

As $k$ approaches $N$, all schemes converge to zero error but differ in convergence smoothness and reliability. Late-stage behavior is best controlled by PSS and AUS through stabilization mechanisms, whereas less structured approaches exhibit mild oscillations as the candidate set shrinks. At very small budgets ($k \le 3$), differences reflect sample representativeness. RS often selects extremes, causing large initial bias and heavy-tailed errors, whereas BRS and QRS impose rank-structured coverage that reduces worst-case error.

In the intermediate regime, adaptivity and stabilization become decisive. QRS exhibits smooth decay due to deterministic coverage but does not exploit residual feedback. PSS improves upon pure coverage by incorporating DT-only dispersion with progressive allocation and tail control. AUS achieves the fastest reduction in mean and maximum error by leveraging residual-informed allocation with EWMA stabilization.

Near budget exhaustion, tail effects dominate. Schemes enforcing rigid structure late in the process can exhibit minor oscillations as feasible locations diminish. PSS mitigates this via explicit late-stage fallback and fading of structure, while AUS maintains stable convergence through estimator smoothing. Although all schemes coincide at $k=N$, convergence reliability and monotonicity differ substantially, with PSS and AUS providing the most predictable behavior.

Table~\ref{tab:reduction} summarizes the average percentage reduction across all measurement budgets in the mean and worst-case offset errors relative to RS, as defined in~(\ref{eq:metrics}). Overall, the results demonstrate effective DT calibration with integrated measurement selection under strict budgets. Combining DT-aware structure, residual-informed adaptivity, and estimator stabilization yields improvements across regimes.
\begin{table}[t!]
\small
\centering
\caption{Average reduction in offset error relative to RS.}
\label{tab:reduction}
\begin{tabular}{lcccc}
\hline
Metric & BRS & QRS & PSS & AUS \\
\hline
Mean (\%) & 11.22 & 19.21 & 26.15 & 72.05 \\
Max (\%)  & 20.00 & 42.64 & 62.29 & 91.04 \\
\hline
\end{tabular}
\end{table}

\section{Conclusion}
This paper studied active calibration of physics-based RF DTs under strict measurement-budget constraints. A site-specific DT pipeline based on semantic geospatial data, 3D scene construction, and deterministic ray tracing was calibrated using limited real radio measurements. By casting calibration as an online learning problem, we examined measurement-selection strategies ranging from unstructured baselines to progressive structure-aware and fully adaptive, uncertainty-aware designs. Adaptive and structure-aware schemes achieved the fastest error decay and smallest worst-case deviations, particularly in the low-budget regime where measurement efficiency is critical. These results highlight the importance of active RF DT calibration for maximizing measurement efficiency and enabling reliable deployment of physics-based DTs under limited measurement budgets.

\bibliographystyle{IEEEtran}
\bibliography{references}

\end{document}